%% file: paper.tex
\documentclass[9pt,journal,compsoc,letter]{IEEEtran}

\usepackage[table,xcdraw]{xcolor}
\usepackage{xfrac}
\usepackage{booktabs}
\usepackage{subfigure}
\usepackage{comment}
\usepackage{enumitem}
\usepackage{caption}
\DeclareCaptionFont{9pt}{\fontsize{9pt}{9pt}\selectfont}
\usepackage[pass]{geometry}
\usepackage{fancyhdr}
\usepackage[normalem]{ulem}
\usepackage{color}
\usepackage{soul}

\usepackage{url}

\usepackage{xspace}
\usepackage{hhline}
\usepackage{makecell}
\usepackage[utf8]{inputenc}
\usepackage[english]{babel}
\usepackage{amsmath}
\usepackage{tcolorbox}
\usepackage[sort,nocompress]{cite}
\usepackage{lipsum}
\usepackage{xfrac}

\newcommand{\name}{X-ray\xspace}
\newcommand{\sixf}{$\text{6F}^\text{2}$\xspace}

\newcommand{\wordline}{WL\xspace}

\newcommand{\bitline}{BL\xspace}
\newcommand{\bitlines}{BLs\xspace}
\newcommand{\senseamp}{SA\xspace}
\newcommand{\senseamps}{SAs\xspace}
\newcommand{\vdd}{$\text{V}_\text{dd}$\xspace}
\newcommand{\effc}{$\text{Eff}_\text{charge}$\xspace}
\newcommand{\effd}{$\text{Eff}_\text{discharge}$\xspace}

\newcommand{\revision}[1]{{\color{black}#1}}

\begin{document}

\title{\name: Discovering DRAM Internal Structure and Error Characteristics by Issuing Memory Commands}
\author{
Hwayong~Nam$^{\dag}$, Seungmin~Baek$^{\dag}$, Minbok~Wi$^{\dag}$, Michael~Jaemin~Kim$^{\dag}$, Jaehyun~Park$^{\dag}$,
Chihun~Song$^{\ddag}$, Nam~Sung~Kim$^{\ddag}$, and ~Jung~Ho~Ahn$^{\dag}$
\\ 
$^{\dag}$Seoul National University, $^{\ddag}$ University of Illinois at Urbana-Champaign
}

\markboth{}%
{}
\IEEEtitleabstractindextext{

\begin{abstract}

The demand for accurate information about the internal structure and characteristics of dynamic random-access memory (DRAM) has been on the rise.
Recent studies have explored the structure and characteristics of DRAM to improve processing in memory, enhance reliability, and mitigate a vulnerability known as rowhammer.
However, DRAM manufacturers only disclose limited information through official documents, making it difficult to find specific information about actual DRAM devices.

This paper presents reliable findings on the internal structure and characteristics of DRAM using activate-induced bitflips (AIBs), retention time test, and row-copy operation.
While previous studies have attempted to understand the internal behaviors of DRAM devices, they have only shown results without identifying the causes or have analyzed DRAM modules rather than individual chips.
We first uncover the size, structure, and operation of DRAM subarrays and verify our findings on the characteristics of DRAM. 
Then, we correct misunderstood information related to AIBs and demonstrate experimental results supporting the cause of rowhammer.
We expect that the information we uncover about the structure, behavior, and characteristics of DRAM will help future DRAM research.
\end{abstract}}

\maketitle

\IEEEdisplaynontitleabstractindextext

\IEEEpeerreviewmaketitle

\input{1_introduction}

\input{2_background}
\input{3_contribution-1}

\input{4_contribution-2}

\input{5_conclusion}

\section*{Acknowledgments}
\revision{This work was in part supported by 
the IITP
grant funded by the Korea government (MSIT) (No. 2021-0-00863) and PRISM, one of the seven centers in JUMP 2.0, a Semiconductor Research Corporation (SRC) program sponsored by DARPA.
The EDA tool was supported by the IC Design Education Center, Korea.
}

\bibliographystyle{IEEEtranS}
\bibliography{ref}

\end{document}

%% file: 1_introduction.tex
\section{Introduction}
\label{sec:DRAM organization and operation}

The recent trend of exacerbating soft and hard error rates, the advent of processing in/using memory, and the discovery of the ever-worsened rowhammer vulnerability~\cite{isca-2020-revisit} have made it more important than ever to deeply understand the internal structure and error characteristics of DRAM.
DRAM is a complex technology with decades of optimizations, and the design choices made by each DRAM manufacturer are proprietary.
Prior studies have used creative reverse-engineering methodologies to disclose partial information about the internal structure and characteristics of DRAM~\cite{isca-2020-revisit, micro-2021-deeper}.
However, we have found some of this information to be misleading, outdated, or limited to a certain type of DRAM.

\revision{
This paper provides a comprehensive study of internal structures and error characteristics of DDR4 and HBM2, utilizing the three known reverse-engineering techniques.
We present new observations of previously undefined behaviors or structures such as asymmetric subarray size, coupled row, and coupled edge subarrays (\S\ref{sec:3_discovering_DRAM_structure}).
Moreover, we push the state-of-the-art understanding of the data-, cell structure-, and chip-dependent DRAM error characteristics, while also clarifying the common pitfall in the interpretations of experimental results (\S\ref{sec: Investigate DRAM characteristics}).
We expect our study to guide future DRAM reliability studies to better fit the real-world DRAM, such as devising a worst-case error pattern aware rowhammer protection mechanism or chip-variation aware side-band/on-die ECC.
}

\section{DRAM Organization and Structure}
\label{sec:2_background}
\begin{figure}[!tb]
  \center
  \vspace{0in}
  \includegraphics[width=1\columnwidth]{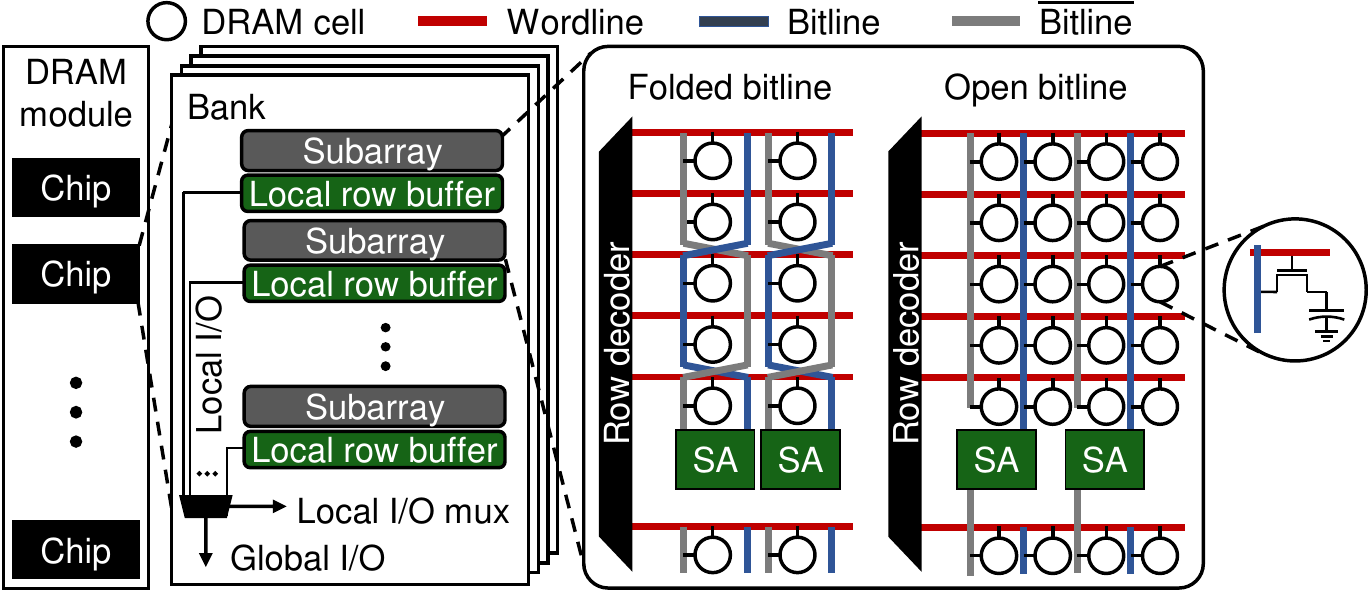}
  \vspace{-0.2in}
  \caption{
    DRAM organization and structure. 
  }
  \vspace{-0.2in}
  \label{fig:DRAM_background}
\end{figure}

A DRAM module is hierarchically organized, from top to bottom, ranks, chips, banks, subarrays, and cells (see Figure~\ref{fig:DRAM_background}).
Each cell consists of a capacitor and access transistor, indexed by row/column address via the wordline (\wordline) and bitline (\bitline), respectively.
A row decoder enables a \wordline, which turns on the access transistors that connect the cell capacitors to the sense amplifiers (\senseamp) via \bitlines.
\senseamp is a cross-coupled inverter that senses and amplifies the differential signal of the \bitline and temporarily stores the value of the DRAM cell.
A subarray can have either an open or folded bitline structure, depending on whether a single \senseamp is connected to both the upper and lower \bitlines (open) or not (folded).
In the open \bitline case, half of the subarray shares the \senseamp with the upper subarray and the other half with the lower subarray.

\begin{figure*}[!tb]
  \center
  \vspace{0in}
  \includegraphics[width=1\textwidth]{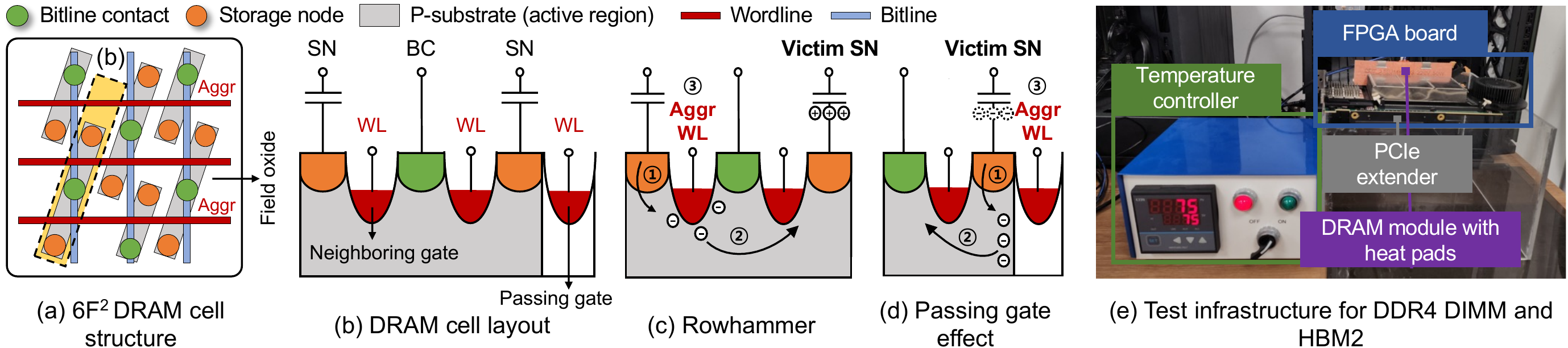}
  \vspace{-0.16in}
  \caption{
    \sixf DRAM cell structure and layout, mechanisms of activate-induced bitflips (AIBs), and test infrastructure of \name.
  }
  \vspace{-0.12in}
  \label{fig:bitflip}
\end{figure*}

Recently, DRAM devices are primarily designed using a \sixf structure~\cite{arxiv-2023-dsac} for higher cell density.
The \sixf DRAM cell adopts a saddle-fin transistor structure with a buried \wordline and has a capacitor connected to a storage node (SN), while a \bitline is connected to a bitline contact (BC) (see Figure~\ref{fig:bitflip}(a) and (b)).
Due to the structure of the \sixf, two cells sharing the same active region belong to different rows.
Moreover, for each row, half of the cells share the active region with the upper row's cells and the other half with the lower row's cells.

As for the DRAM operation, the \senseamps and \bitlines are initially precharged to a voltage of \vdd/2.
To access data, the host sends an activate (\texttt{ACT}) command to connect and charge-share a row of DRAM cells with \bitlines by enabling a \wordline.
Charge-sharing causes a small deviation in the voltage level of the \bitline.
The \senseamp amplifies it to full \vdd or 0.
When DRAM receives a read or write command, the sensed or to-be-written data pass through the local and global I/O, equipped with temporary buffers on its path (e.g., global dataline \senseamp).
After completing read or write operations, the host sends a precharge (\texttt{PRE}) command to disable the activated row's \wordline, disconnecting the \bitlines and cells. 
Prior to the precharge, the voltage level of the cell must be restored to full \vdd or 0.
The required time to issue the \texttt{PRE} command after the \texttt{ACT} command is \texttt{tRAS}.
Moreover, after issuing of \texttt{PRE} command, the \senseamps and \bitlines require \texttt{tRP} time to restore a voltage of {\vdd}/{2}.

%% file: 2_background.tex
\section{DRAM Reverse-engineering Techniques}
\label{sec:2_tech_and_exp_setup}
We leverage three reverse-engineering techniques to identify and verify the DRAM internal structures and characteristics: 1) causing activate-induced bitflips, 2) executing row-copy operation, and 3) inducing retention errors.

\textbf{Activate-induced bitflips (AIBs)} \revision{are a DRAM error phenomenon in which the victim cell experiences bitflips when the neighboring aggressor is activated with certain conditions.}
There are two types of AIBs: 1) rowhammer and 2) passing gate effect~\cite{arxiv-2023-dsac}.
\emph{Rowhammer} is a phenomenon in which repetitive activation of \revision{the aggressor} causes bitflip in the opposite victim cell that shares the active region (see Figure~\ref{fig:bitflip}(c)).
When an \revision{aggressor's} \wordline is activated, \textcircled{\scriptsize{1}} electrons accumulate around the \wordline due to a channel inversion. 
Upon deactivation, \textcircled{\scriptsize{2}} the accumulated electrons are spread out and some are injected into the opposite victim cell.
\emph{Passing gate effect} conversely refers to the occurrence of a bitflip in a victim cell that is separated from the aggressor by a field oxide, due to the repetitive activation of the \wordline over an extended period of time (see Figure~\ref{fig:bitflip}(d)). 
When the \revision{aggressor's} \wordline is activated, \textcircled{\scriptsize{1}} electrons are continuously attracted from the \revision{victim} cell toward the passing gate. 
After the row is precharged, \textcircled{\scriptsize{2}} the electrons are spread out and \revision{some are injected into the active region, instead of returning to the victim cell.}
\revision{
In both cases of rowhammer and passing gate effect, a repetition of such process results in a bitflip in the victim cell.
As both are the processes of victim cells acquiring or losing electrons, their likelihoods are affected by the data written to the victim cells.
Besides, it has been reported that rowhammer is more sensitive to the number of activation, whereas the passing gate effect is more affected by the activated time~\cite{arxiv-2023-dsac}.
}

\textbf{Row-copy}~\cite{micro-2013-rowclone, micro-2019-compdram} is an out-of-specification in-memory operation that copies the value of one row to another row within the same subarray using charge-sharing through the \bitline.
First, a source row is activated. After \texttt{tRAS}, the row is precharged. 
However, if the destination row is activated soon enough, the \bitline will not be fully precharged to \vdd/2. 
Because the capacitance of \bitline is much larger than the cell, the source row values are effectively copied to the destination row.
Leveraging the fact that \bitline charge-sharing is the fundamental cause of the successful row-copy operation, we not only identify the size of the subarray, but also identify the type of subarray structure (open or folded bitline) for each tested DRAM.

\textbf{Retention time test} exploits the fact that DRAM cells naturally leak charge over time, which leads to retention failure unless periodically refreshed.
The retention time of a cell is the length of time before it loses data.
Exploiting the fact that leakage occurs from a charged state to a discharged state, we execute retention time testing to identify the true- and anti-cells.
While some cells store the value 1 as a charged state (true-cell), others store it as a discharged state (anti-cell), which depends on the design choice with the intention of reducing the noise or optimizing the data path from the \senseamps to the I/O~\cite{isca-2013-trueanti}.

\vspace{-0.09in}
\section{Experimental Setup}
\label{sec:2_4_exp_setup}

We modified SoftMC~\cite{hpca-2017-softmc} to execute the three techniques on DDR4 and HBM2 (see Figure~\ref{fig:bitflip}(e)).
We tested 192 DDR4 chips (8Gb $\times$4 chips) from two DRAM manufacturers A (manufactured in 2016 and 2018) and B (manufactured in 2018 and 2021), and two HBM2 cubes (4GB/cube) with undisclosed manufacturers.
DDR4 and HBM2 are controlled at 1.25 ns and 2.50 ns, respectively, using Xilinx Alveo U280 FPGA boards~\cite{alveo}.
We also employed a temperature controller and silicon rubber heaters to regulate the temperature of DRAM. 
We tested the DDR4 DIMMs at a temperature of 75\textdegree C and HBM2 at room temperature as we could not regulate the temperature of HBM2.

We emphasize that correctly interpreting \emph{the physical address to DRAM address remapping} is essential in acquiring correct information from the reverse-engineering techniques.
For example, row addresses are remapped in the row decoder, inverted at a registered clock driver (RCD) chip for half of the DRAM chips~\cite{jedec-ddr4-rcd}, and DQ pins are shuffled on a DIMM~\cite{jedec-ddr4-rdimm}.
We will later clarify that some misconceptions of prior studies stem from the misinterpreted address remapping. 

%% file: 3_contribution-1.tex
\vspace{-0.09in}
\section{Discovering DRAM structures}
\label{sec:3_discovering_DRAM_structure}

\noindent
\textbf{Subarray sizes} of each DRAM chip is verified using all three reverse-engineering techniques and cross-checked. 
First, because two different subarrays are separated by \senseamps, we look for row address boundaries where AIB occurs from only one aggressor row. 
Also, considering that only half of the cells share \senseamps with the upper/lower subarray in the case of open bitline structure, we look for row address boundaries where row-copy starts to work only for half the cells.
Moreover, in the case of the manufacturer (mfr.) B, we can also look for boundaries of true- or anti-cells as it is known that a single subarray consists of only one type of cells~\cite{isca-2013-trueanti}.

As opposed to the common understanding of the size of a subarray, we discovered that the size of subarrays is not a power of 2 and also varies within even a single chip.
The mfr. A's DDR4 chips manufactured in 2016 have a repeated pattern of 11 subarrays of 640 rows and two subarrays of 576 rows (a total of 8192).
In the case of chips made in 2018, a pattern of 
four subarrays of 832 rows and one subarray of 768 rows (total of 4096) is repeated.
By contrast, the mfr. B's DDR4 chips have a pattern of two subarrays with 688 rows and one subarray with 672 rows (a total of 2048).
Lastly, HBM2 chips have a pattern that repeats in units of 4096 rows, with each pattern consisting of four subarrays of 832 rows and one subarray of 768 rows.
We argue that the varying size of the subarray is a compromise between increasing timing parameters and higher cell density when the cell per \bitline (subarray size) increases.
This concurs with the fact that the size of the subarray is on an increasing trend, following the DRAM process scaling.

\begin{figure}[!tb]
  \center
  \vspace{0in}
  \includegraphics[width=1\columnwidth]{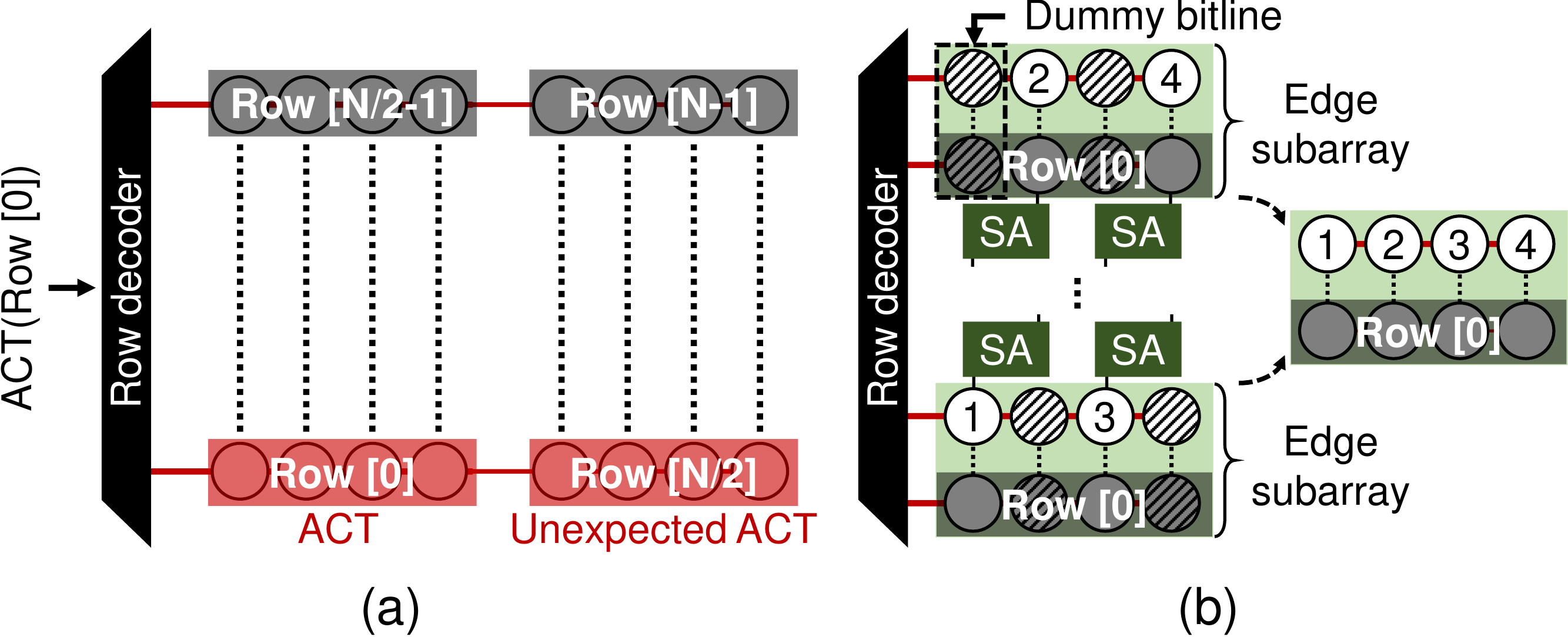}
  \vspace{-0.16in}
  \caption{
    Characteristics of subarray structures: (a) Activating a row could incur activating its coupled row. (b) An edge subarray physically consists of a pair of subarrays, each having dummy bitlines.
  }
  \vspace{-0.2in}
  \label{fig:subarray_structure}
\end{figure}

\vspace{-0.04in}
\begin{tcolorbox}[boxsep=0pt,left=3pt,right=3pt,arc=0pt]
\emph{\textbf{Observation-1:}}
The subarray sizes are not power of 2, and are different across different generations and within a chip. 
\end{tcolorbox}
\vspace{-0.05in}

\textbf{Subarray structure types} are verified by exploiting the row-copy operation.
Because adjacent subarrays share half of the \senseamps for open bitline structure and none for folded bitline structure, checking the copied data from the row-copy operation across different subarrays allows us to distinguish them.
In the case of folded bitline, none of data is changed, whereas half of the row is copied in the case of an open bitline structure.

All the tested DRAM chips have an open bitline structure.
However, while the row-copy on mfr. B's DDR4 resulted in half of the row being copied as is, DDR4 of mfr. 
A and HBM2 resulted in the copied values being inverted.
While mfr. B's DDR4 consists of both true-/anti-cells, HBM2 consists only of true-cells.
We believe that this is due to the design choice in the datapath, where the latter would employ a MUX to choose one of the differential local I/O signals.

\textbf{Coupled rows activation} is also identified for a portion of $\times$4 DRAM chips; when one row is activated, its coupled row is simultaneously activated.
Both the row-copy operation and the AIBs indicate that when a row is activated (e.g., $i^{th}$ row), its coupled row (e.g., $(i+N_{row}/2)^{th}$ row) is also activated, where $N_{row}$ denotes the total number of rows in a bank.
Such behavior was exhibited on mfr. A's $\times$4 DDR4 chips and HBM2.
Thus, we speculate that this is a result of optimization to reduce the number of row address decoders or to maintain a uniform internal DRAM cell structure between the chips with different I/O widths (i.e., $\times$8 and $\times$4 chips).
This behavior can serve as another vulnerability regarding AIBs unless the host is aware of this pairing and applies proper mitigation to both the victim row and its coupled row.

\vspace{-0.05in}
\begin{tcolorbox}[boxsep=0pt,left=3pt,right=3pt,arc=0pt]
\emph{\textbf{Observation-2:}}
For some DRAM chips, activating a row can result in the unintended activation of the coupled row.
\end{tcolorbox}
\vspace{-0.05in}

\textbf{Edge subarrays} of the open bitline structure were also identified to work in tandem to create a single full subarray.
For some tested DRAM chips, when the row-copy operation was executed for $0^{th}$ row as a source and $(N_{row}/2-1)^{th}$ row as a destination, half of the cells were copied despite the large difference in the row address values.
Because the $0^{th}$ row and the $(N_{row}/2-1)^{th}$ row belong to the subarrays of the bottom and top edge, respectively, we speculate that these two subarrays work together as a single subarray.
Similarly, the DDR4 from mfr. A manufactured in 2016 and 2018 have edge subarrays at every $N_{row}/8$ and $N_{row}/4$ boundary, respectively.
That of the DDR4 from mfr. B and manufacturer-unspecified HBM2 was $N_{row}/4$ and $N_{row}/2$, respectively.
Such a structure of the edge subarray is reasonable considering that only half of the cells are connected to \senseamps on either side of the edge (see Figure~\ref{fig:subarray_structure}(b)).
Also, this is an undefined behavior for the host and thus could be a source of vulnerability.

\vspace{-0.05in}
\begin{tcolorbox}[boxsep=0pt,left=3pt,right=3pt,arc=0pt]
\emph{\textbf{Observation-3:}}
For some DRAM chips with open bitline structure, two edge subarrays work in tandem to create a single full subarray.
\end{tcolorbox}
\vspace{-0.2in}

%% file: 4_contribution-2.tex
\section{Investigating DRAM AIB characteristics}
\label{sec: Investigate DRAM characteristics}

\begin{figure}[!tb]
  \center
  \includegraphics[width=1\columnwidth]{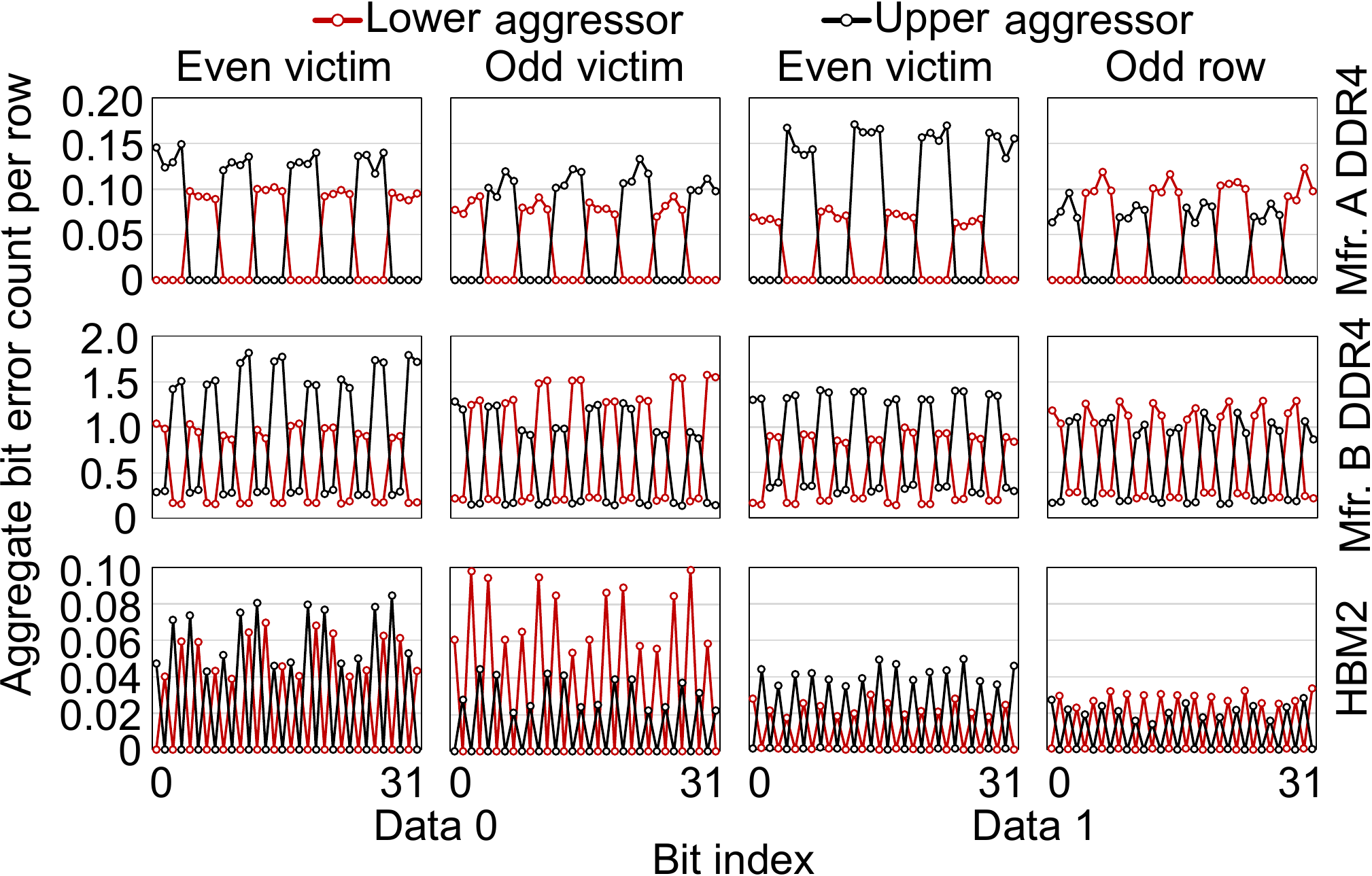}
  \vspace{-0.16in}
  \caption{
    Aggregate bit error count from AIB of DDR4 and HBM2. Considering the recurring pattern, we only show the bit index up to 32.
  }
  \vspace{-0.1in}
  \label{fig:6f2_graph}
\end{figure}

\begin{figure}[!tb]
  \center
  \vspace{0in}
  \includegraphics[width=1\columnwidth]{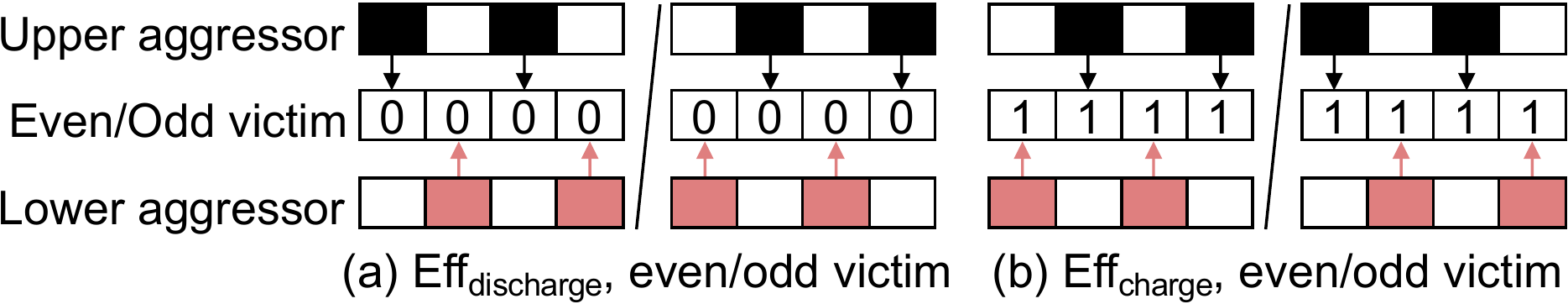}
  \vspace{-0.16in}
  \caption{
    Exemplar \effc and \effd for HBM2.
  }
  \vspace{-0.2in}
  \label{fig:effective_aggressor_pattern}
\end{figure}

In this section, we demonstrate for the first time that the worst-case AIB aggressor pattern is defined by the \sixf DRAM structure as well as the data dependence of rowhammer and passing gate effect.
Also, we present the intra-/inter-chip AIB variation and clarify the commonly misperceived AIB characteristic.

First, we observed that when each DRAM cell is AIB victimized, it exhibited a particularly more vulnerable side of aggressor depending on the written data and its position.
For example, half of the cells in a row were more vulnerable to the upper aggressor when in a charged state and to the lower aggressor when in a discharged state (see Figure~\ref{fig:6f2_graph}).
More specifically, there existed a pattern of such correlation for each cell depending on the cell index and whether they were in an even/odd row.
We believe that such a pattern originates from the \sixf structure, where each cell has a neighboring gate (rowhammer vulnerable) on one side and a passing gate (passing gate effect vulnerable) on the other.
Because both rowhammer and passing gate effect are affected by the written data of the victim cell, the correlation between the data and the vulnerable side can also be explained.
While \sixf should expose a pattern repeating in a cell index of 2, some chips in Figure~\ref{fig:6f2_graph} demonstrate a pattern of 8 and 4 due to the difference in serialization inside the DRAM chip.

\begin{figure}[!tb]
  \center
  \vspace{0in}
  \includegraphics[width=0.96\columnwidth]{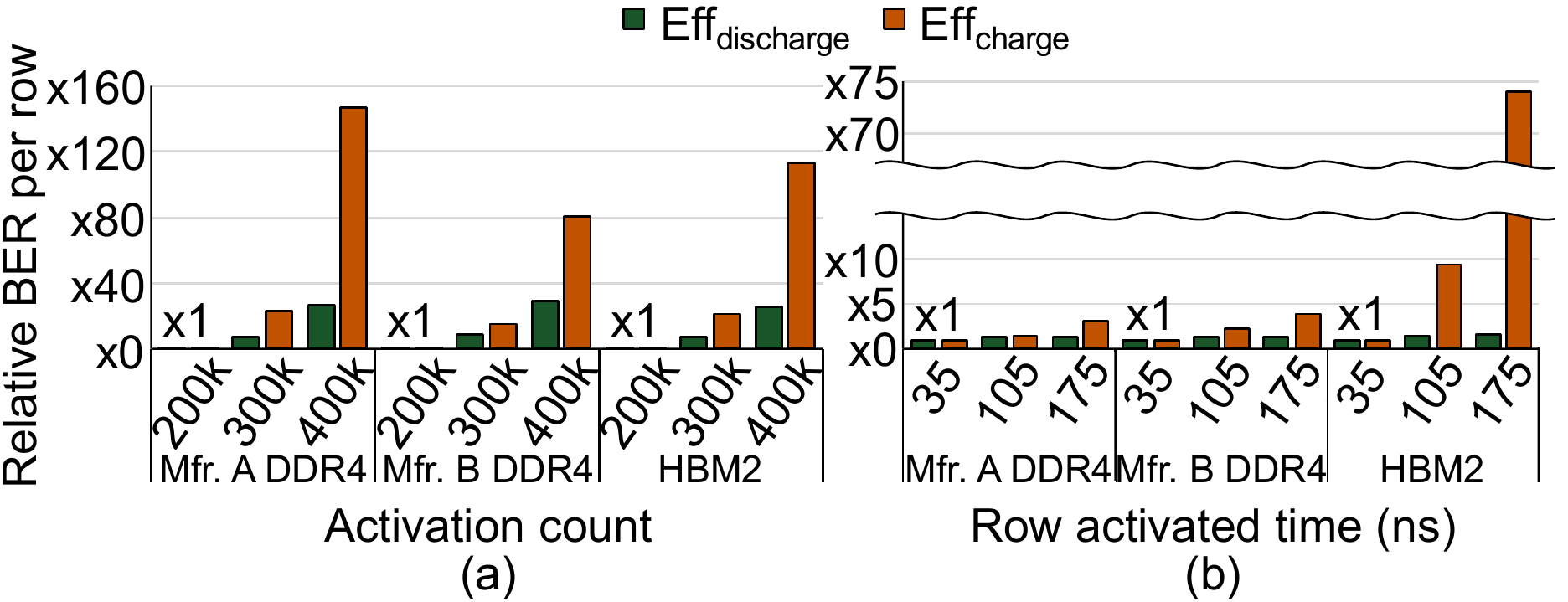}
  \vspace{-0.08in}
  \caption{
    Relative BER (bit error rate) per row of \effc and \effd when we vary activation counts and row activated time for the DDR4 and HBM2 devices.
  }
  \vspace{-0.15in}
  \label{fig:activate_bitflips}
\end{figure}

To further investigate the correlation of our observation with rowhammer and passing gate effect, we conducted a sensitivity study on the \emph{activation count} (rowhammer sensitive) and \emph{row activated time} (i.e., duration of a row staying active, which is passing-gate sensitive).
We first categorized two types of AIB aggressor patterns; i) aggressor patterns that are effective when the victim data is in a charged state (\effc) and ii) in a discharged state (\effd).
As Figure~\ref{fig:effective_aggressor_pattern} illustrates, both patterns have interleaving upper/lower sides depending on the victim data, whether the victim is an even/odd row, and the bit index.
Each pattern should be correlated to the rowhammer or passing gate effect, respectively.
Upon these two patterns, we changed the activation count from 200K to 400K with \texttt{tRAS} (35ns) activated time, and changed the activated time from 35ns to 175ns with a fixed activation count of 300K.

First, both patterns of \effc and \effd were highly sensitive to the activation count with \effc reaching up to 146$\times$ increase in bit error rate (BER) per row (see Figure~\ref{fig:activate_bitflips}(a)).
Second, while \effc was also sensitive to the activated time, \effd was relatively less sensitive (see Figure~\ref{fig:activate_bitflips}(b)).
The BER per row increase for \effd was limited to less than 1.52$\times$. 
Based on these results, we conclude that \effc, which is sensitive to activate time, is the result of the passing gate effect and \effd, which is \emph{only} sensitive to activation count, is the result of the rowhammer.
This contradicts the prior explanation~\cite{arxiv-2023-dsac} that states that rowhammer is effective on the charged state victim and vice versa for the passing gate, which requires further investigation in the future.

\vspace{-0.05in}
\begin{tcolorbox}[boxsep=0pt,left=3pt,right=3pt,arc=0pt]
\emph{\textbf{Observation-4:}}
The worst-case AIB aggressor pattern is defined by the unique \sixf DRAM cell structure and data dependence of rowhammer and passing gate effect.
\end{tcolorbox}
\vspace{-0.05in}

\begin{figure}[!tb]
  \center
  \vspace{0in}
  \includegraphics[width=1\columnwidth]{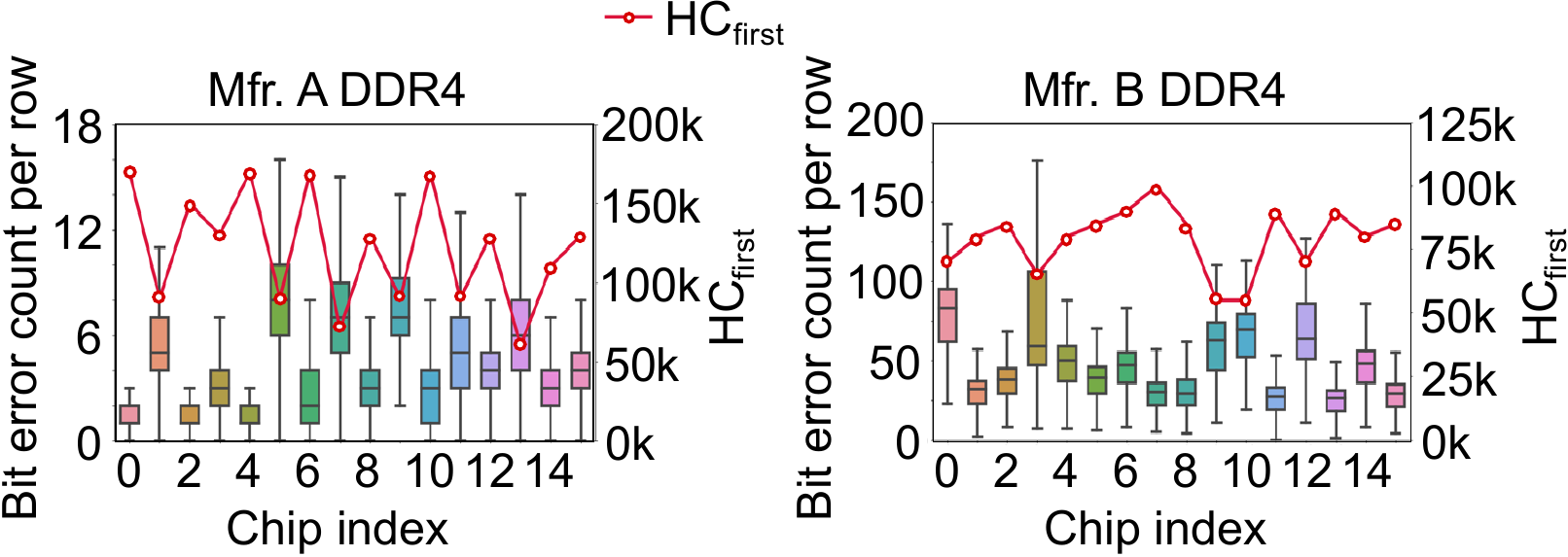}
  \vspace{-0.16in}
  \caption{
    The box and whisker plots of the bit error count per row for each chip of DDR4 and HBM2 and the line graphs of $\text{HC}_\text{first}$ for each chip of DDR4.
  }
  \vspace{-0.12in}
  \label{fig:variation}
\end{figure}

We also recognized that the AIB characteristics exhibit relatively small intra-chip and large inter-chip variations.
Figure~\ref{fig:variation} demonstrates the box and whisker plot of the bit error count of each row for a fixed 300K number of activations, and the line plot of $\text{HC}_\text{first}$ (the lowest activation count that incurs first bitflip anywhere on a chip).
We identify that while there were some variations inside a single chip (intra-chip), chip-to-chip variation was more pronounced (inter-chip).
This indicates that an AIB vulnerable chip can determine the overall vulnerability of the whole DRAM module.

\vspace{-0.05in}
\begin{tcolorbox}[boxsep=0pt,left=3pt,right=3pt,arc=0pt]
\emph{\textbf{Observation-5:}}
AIB exhibits significant inter-chip variation while demonstrating a relatively restricted distribution within a single chip.
\end{tcolorbox}
\vspace{-0.05in}

Finally, we verified that the direct non-adjacent rowhammer (AIB) effect\footnote{A phenomenon where frequently activating $N^{th}$ row can \emph{directly} affect not only distance 1 (i.e., $N\!\pm\!1^{th}$ rows) but also distance 3, 5 and further away rows~\cite{isca-2020-revisit}.}~\cite{isca-2020-revisit} and half row~\cite{so-2020-suscep} are misguided interpretations which can be clarified when the RCD address inversion~\cite{jedec-ddr4-rcd} is properly taken into account.
Half of the DRAM chips in a registered DIMM experience address inversion at the RCD chip~\cite{jedec-ddr4-rcd}.
The direct non-adjacent AIB effect was able to be reproduced, yet only when such an address inversion by RCD chips was neglected.
This concurs with the prior study that could not reproduce the direct non-adjacent AIB effect~\cite{github-2021-halfdouble}.
The same was true for the half row observation~\cite{so-2020-suscep}.
We believe that our clarification further highlights the complexity and difficulty of correctly reverse-engineering the DRAM internals.

%% file: 5_conclusion.tex
\section{Future work and conclusion}
\label{sec:5_future_work_conclusion}

We have reliably revealed the DRAM's internal structure and activate-induced bitflip (AIB) characteristics through AIB tests, row-copy operation, and retention tests using commercial DRAM devices.
We discovered the previously undisclosed subarray structure and behaviors, and also the worst-case AIB aggressor pattern that is determined by the \sixf structure and data dependence of rowhammer and passing gate effect.
We also clarified the common misconceptions from prior DRAM studies such as direct non-adjacent AIB effect.
We anticipate our new observations and clarifications, as well as the experimental methodology itself, to facilitate the future DRAM research.